\documentclass[a4paper,11pt]{article}
\usepackage{pos}
\usepackage{color}
\usepackage{slashed}
\newcommand{\DNS}[1]{{\color{orange}#1}}
\newcommand{\MeV}{\mathrm{MeV}}
\newcommand{\GeV}{\mathrm{GeV}}
\newcommand{\fm}{\mathrm{fm}}
\newcommand{\tsep}{t_\text{sep}}
\newcommand{\tins}{t_\text{ins}}
\newcommand{\la}{\langle}
\newcommand{\ra}{\rangle}
\newcommand{\Dslash}{\slashed{D}}

\title{Nucleon electromagnetic form factors at large momentum from Lattice QCD}

\author*[a]{S.~Syritsyn}
\emailAdd{sergey.syritsyn@stonybrook.edu}
\author[b]{M.~Engelhardt}
\author[c,d,e]{S.~Krieg}
\author[f]{J.~Negele}
\author[f]{A.~Pochinsky}

\affiliation[a]{Department of Physics \& Astronomy, Stony Brook University,
  1 Nicolls Rd, Stony Brook, NY 11733, USA}

\affiliation[b]{Department of Physics, New Mexico State University,
  Las Cruces, NM 88003, USA}

\affiliation[c]{J\"ulich Supercomputing Centre,
  Forschungszentrum J\"ulich, 52425 J\"ulich, Germany}

\affiliation[d]{Helmholtz-Institut f\"ur Strahlen- und Kernphysik,
  Universit\"at Bonn, 53115 Bonn, Germany}

\affiliation[e]{Center for Advanced Simulation and Analytics (CASA),
  Forschungszentrum J\"ulich, 52425 J\"ulich, Germany}

\affiliation[f]{Center for Theoretical Physics, Massachusetts Institute of Technology,
  77 Massachusetts Ave., Cambridge, MA 02139, USA}

\abstract{Proton and neutron electric and magnetic form factors are the primary characteristics
of their spatial structure and have been studied extensively over the past half-century.
At large values of the momentum transfer $Q^2$ they should reveal transition from
nonperturbative to perturbative QCD dynamics as well as effects of quark orbital angular momenta and
diquark correlations.
Currently, these form factors are being measured at JLab at momentum transfer up to $Q^2=18$
GeV$^2$ for the proton and up to 14 GeV$^2$ for the neutron.
We report preliminary results of our lattice calculations of these form factors,
including $G_E$ and $G_M$ nucleon form factors with momenta up to $Q^2=8$ GeV$^2$, pion masses
down to the almost-physical $m_\pi$=170 MeV, several lattice spacings down to $a=0.073$ fm, and
high $O(10^5)$ statistics.
Specifically, we study individual form factors, the $G_E/G_M$ ratios, and flavor dependence of
contributions to the form factors.
We observe qualitative agreement of our ab initio theory calculations with experiment.
Comparison of our calculations and upcoming JLab experimental results will be an important test
of nonperturbative QCD methods in the almost-perturbative regime.}

\FullConference{The 41st International Symposium on Lattice Field Theory (LATTICE2024)\\
 28 July - 3 August 2024\\
Liverpool, UK\\}

\begin{document}
\maketitle

\section{Introduction}


%
%

Nucleon electromagnetic form factors $G_{Ep,n}$, $G_{Mp,n}(Q^2)$ at high momentum transfer
$Q^2\approx5\ldots10\,\mathrm{GeV}^2$ describe short-scale spatial distributions of electric
charge and magnetization and are crucial to our understanding of nucleon structure and improving
nucleon models.
In general, they are defined as nucleon matrix elements of the vector current,
\begin{gather}
\label{eqn:formfacdef}
\langle N(p', S')|\bar q\gamma^\mu q| N(p,S)\rangle = \bar u_{p',S'} \Big[
F_1(Q^2) \gamma_\mu + F_2(Q^2)\frac{i\sigma^{\mu\nu}(p'-p)_\nu}{2m_N}\Big] u_{p,S}\\
\nonumber\text{ and } 
G_E(Q^2) = F_1(Q^2) - \frac{Q^2}{4M^2} F_2(Q^2)\,,\quad
G_M(Q^2) = F_1(Q^2) + F_2(Q^2)\,,
\end{gather}

for spacelike $Q^2 = - (p' - p)^2 \ge0$.
Various models such as vector meson dominance, chiral solitons, pion cloud, and relativistic constituent
quarks have been tried to predict form factor behavior at large $Q^2$. 
While these models generally describe available form factor data, their predictions differ
outside of experimentally explored range of momenta~\cite{Punjabi:2015bba}.
It has also been shown using Dyson-Schwinger and Faddeev equations that incorporating diquark
correlations is important for understanding nucleon electromagnetic structure at high momentum
transfer~\cite{Cui:2020rmu}.
For example, quark correlations in the Faddeev's amplitude of the proton determine if there is indeed a
peculiar zero-crossing in the electric Sachs form factor $G_{Ep}$ around $Q^2\approx 8.5\GeV^2$,
and thus can be inferred from experimental or nonperturbative lattice form factor data.
The experimental program to determine nucleon form factors up to $Q^2\approx 18\,\mathrm{GeV}^2$
is well underway~\cite{jlabHallA_HRS_gmp17GeV,jlabHallA_SBS_gep15GeV,jlabHallA_SBS_gen10GeV,
jlabHallA_SBS_gmn18GeV,jlabHallB_CLAS12_gmn14GeV},
and the first results have been published for the proton magnetic form factor $G_{Mp}(Q^2)$ for $Q^2$ up to
$\approx16\text{ GeV}^2$~\cite{Christy:2021snt}.
Our ongoing lattice calculations of the nucleon form factors are intended to complement these
experimental efforts, and have to be performed with rigorous control of all systematic
uncertainties.

Mainstream studies of nucleon form factors on a lattice are usually limited to the region
$Q^2\lesssim1\ldots2\,\mathrm{GeV}^2$. 
One notable exception is the calculation of the $G_{Ep}/G_{Mp}$ ratio using the Feynman-Hellman
method~\cite{Chambers:2017tuf}.
Precise calculation of nucleon structure involving large momenta $|\vec p|\gtrsim m_N$ are
challenging for several reasons.
With growing energy of the in- and out-states, Monte Carlo fluctuations of the corresponding
correlators increase rapidly~\cite{Lepage:1989hd}, and the signal-to-noise ratio decreases
$\propto\exp\big[-(E_N(\vec p) - \frac32 m_\pi)\tau\big]$ with Euclidean time $\tau$.
At the same time, systematic uncertainties from nucleon excited states have weaker suppression
as the energy gaps $\Delta E(\vec p) = E_{N,\text{exc}}(\vec p) - E_N(\vec p)$ shrink due to
the relativistic dispersion relation.
Both these challenges can be minimized by choosing the Breit frame on a lattice, so that the
initial and final momenta of the nucleon are equal to $|\vec p^{(\prime)}|=\frac12\sqrt{Q^2}$.
Still, the minimal nucleon momentum to achieve $Q_1^2\approx10\,\mathrm{GeV}^2$ is 
$p_1\gtrsim1.6\,\mathrm{GeV}$ and results in reducing the typical excitation energy gap 
$\Delta E_N(0)\approx0.5\,\mathrm{GeV}$ to $E_N(p_1)\approx0.3\,\mathrm{GeV}$, which is
challenging even if pion-nucleon states are not considered.
These problems are compounded in computing quark-disconnected nucleon-current correlators,
which could yield substantial contributions to the nucleon form factors at $Q^2\gtrsim1\,\GeV^2$.
Therefore, very large Monte Carlo statistics combined with rigorous analysis of excited states
are crucial for obtaining credible results.

Over a few years, we have accumulated significant Monte Carlo statistics on nucleon-current
correlators with momentum transfer values $Q^2\lesssim8\text{ GeV}^2$ (and larger in some cases) 
at a range of lattice spacings, as well as quark masses that approach the physical point.
Some of our initial results have been previously reported in 
Refs.~\cite{Syritsyn:2017jrc,Kallidonis:2018cas,Syritsyn:2023pmn}.
More recently, we have also accumulated substantial statistics on the disconnected quark loops.
In this paper, we report and compare these disconnected contributions to the connected
contributions on two ensembles, one of which is very close to the physical pion mass.

\section{Lattice setup}

\begin{figure}
\centering
\includegraphics[width=.47\textwidth]{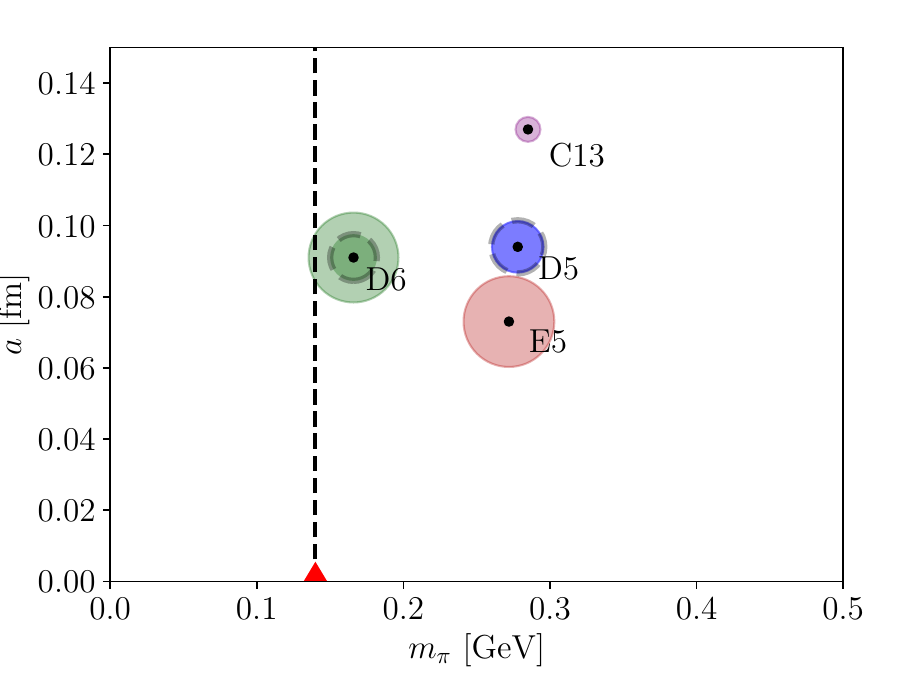}
\caption{Summary of statistics accumulated so far in out study of the form factors.
  The circle area is proportional the number of lattice correlator samples:
  the solid- and dashed-border circles refer to the sample counts for 
  the connected and disconnected contributions, respectively.
  \label{fig:ens_stat}}
\end{figure}

We have performed large-statistics calculations on four ensembles of lattice gauge fields.
We use gauge configurations generated with  $N_f=2+1$ dynamical quarks with isotropic clover-improved
Wilson fermion action by the JLab/W\&M/LANL/MIT groups.
Our ensemble parameters and accumulated statistics are summarized in Tab.~\ref{tab:ens_stat}.
Because discretization effects are potentially significant at large momentum, we use a range of lattice
spacings $a\approx0.07\ldots0.09\,\mathrm{fm}$,  with some earlier calculations performed at
the coarse $a\approx0.127\,\fm$.
While the quark masses should have only limited effect at the scale of $Q^2\gtrsim1\,\GeV^2$,
they may affect energies of excited states contributing to systematic effects, therefore we
study two values of the pion mass, $m_\pi\approx280$ and $170\,\mathrm{MeV}$.
Calculations at the lighter pion mass require substantially more statistics; 
we illustrate the comparison of accumulated data sample counts in Fig.~\ref{fig:ens_stat}.
\begin{table}[ht]
\centering
\caption{Summary of ensembles, form factor kinematic ranges, and statistics.
The columns 5,6,7 show the total number of gauge configurations analyzed, the number of connected
correlator MC samples, and the number of gauge configurations on which disconnected
contributions have been calculated.
\label{tab:ens_stat}}
\begin{tabular}{llrr|rrr|crr}
\hline\hline
ens   & lat & $a$   & $m_\pi$
  & N\text{conf} & $N_\text{stat}^\text{conn}$ & $N_\text{conf}^\text{disc}$
  & $\tsep / a$   & $\tsep$   & $Q_\text{max}^2$ \\
      &     & [fm]  & [MeV]
  & & &
  &               & $[\fm]$      &  $[\GeV^2]$ \\
\hline
\texttt{C13}  & $32^3\times96$   & 0.127   & 285 
  &  210 &  20,160 & --
  & $6\ldots10$ & 1.27 & 8.3  \\
\texttt{D5}   & $32^3\times64$   & 0.094   & 278 
  & 1346 &  86,144 & 1346
  & $6\ldots12$ & 1.13 & 10.9 \\
\texttt{D6}   & $48^3\times96$   & 0.091   & 166 
  & 2040 & 261,120 & 1152
  & $6\ldots12$ & 1.09 & 8.0 \\
\texttt{E5}   & $48^3\times128$  & 0.073   & 272 
  & 2080 & 266,240 & --
  & $7\ldots14$ & 1.02 & 8.0 \\
\hline\hline
\end{tabular}
\end{table}

In order to calculate the nucleon form factors, we evaluate nucleon-current correlators using
the standard sequential propagator method,
\begin{equation}
\label{eqn:c3pt}
C_{N V_q^\mu \bar N}(\vec p', \vec q ; \tsep,\tins) 
  = \sum_{\vec y,\vec z} e^{-i\vec p'\vec y + i\vec q\vec z }\,
    \langle N_{(\vec k)}(\vec y,\tsep)\, [\bar q \gamma^\mu q]_{\vec z,\tins} \, N_{(-\vec k)}(0)\rangle\,,
\end{equation}
where $N_{\pm\vec k}
=\epsilon^{abc}[\tilde{u}_{(\pm\vec k)}^{aT} C\gamma_5 \tilde{d}_{(\pm\vec k)}^b]\tilde{u}_{(\pm\vec k)}^c$ 
is the nucleon interpolating field on a lattice constructed with ``momentum-smeared'' quark
fields $\tilde{q}_{(\pm\vec k)}$, $\vec k \uparrow\uparrow\vec p'$,  to improve their overlap with
the ground states of the boosted in- and out-nucleon~\cite{Bali:2016lva}.
Wick contractions of lattice quark fields generate two types of diagrams: quark-connected and
quark-disconnected. 
At momenta $Q^2\lesssim 1\,\GeV^2$, the form factors are dominated by the connected
contributions discussed in Sec~\ref{sec:conn}.
The quark-disconnected contributions are calculated as correlators between nucleon 2-point Wick contractions
and the quarks loops and are discussed in Sec.~\ref{sec:disc}.
Although the contributions of the latter were found small ($\lesssim1\%$)
at $Q^2\lesssim 1\,\GeV^2$~\cite{Green:2015wqa}, they have to be explored at higher momenta as
they can contribute to systematic uncertainty.

\begin{figure}
\centering
\includegraphics[width=.47\textwidth]{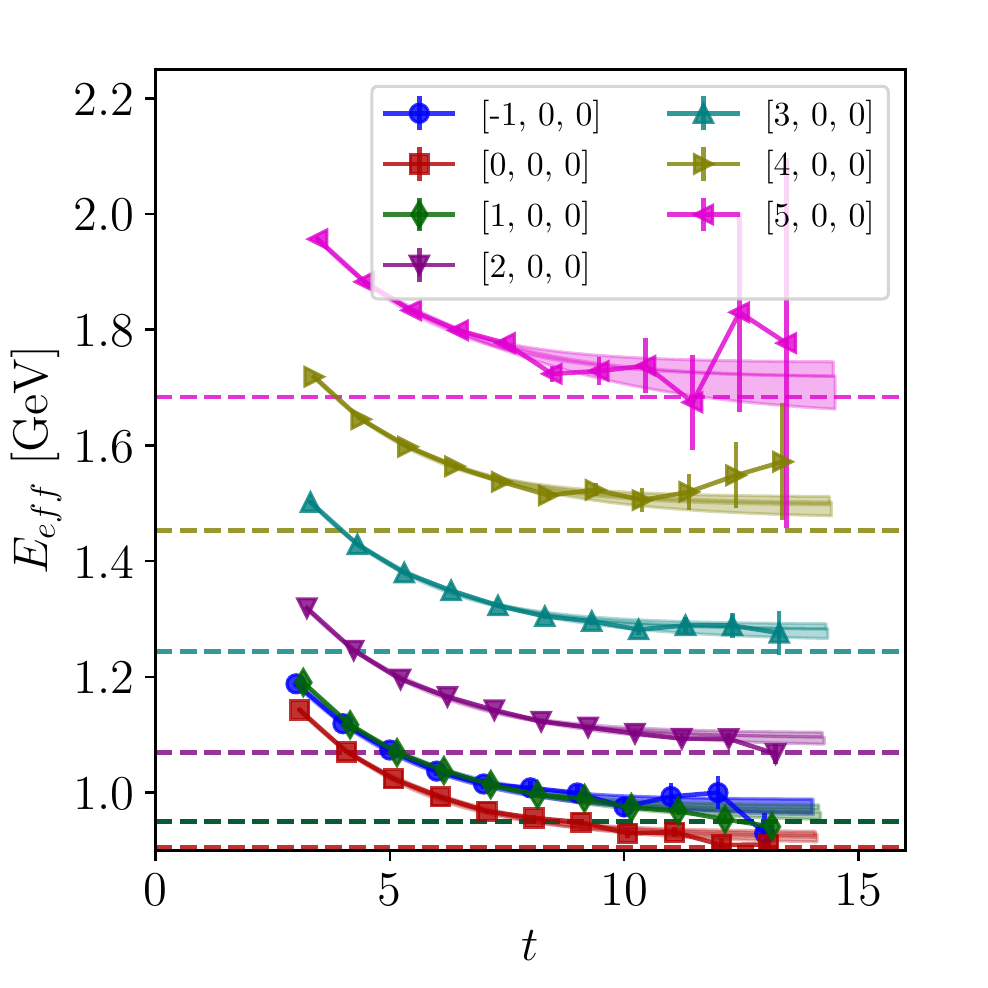}~
\includegraphics[width=.47\textwidth]{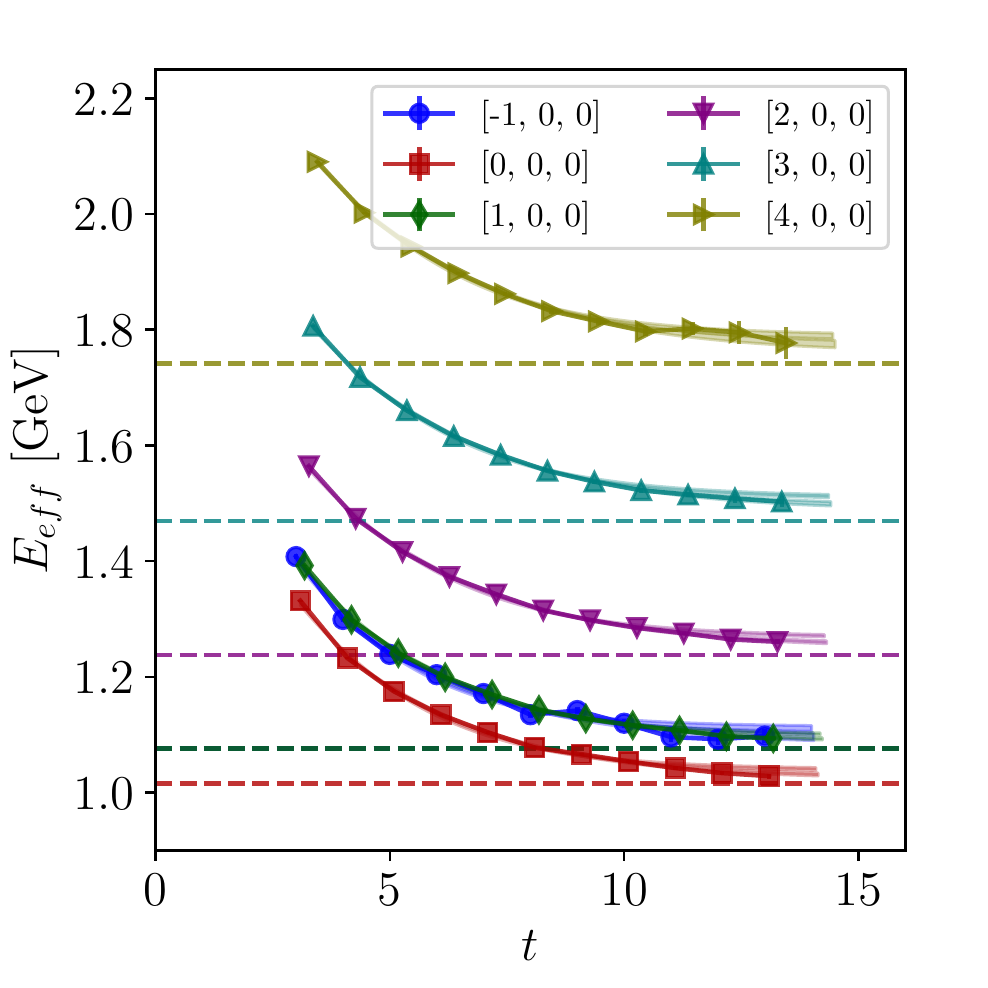}\\
\includegraphics[width=.47\textwidth]{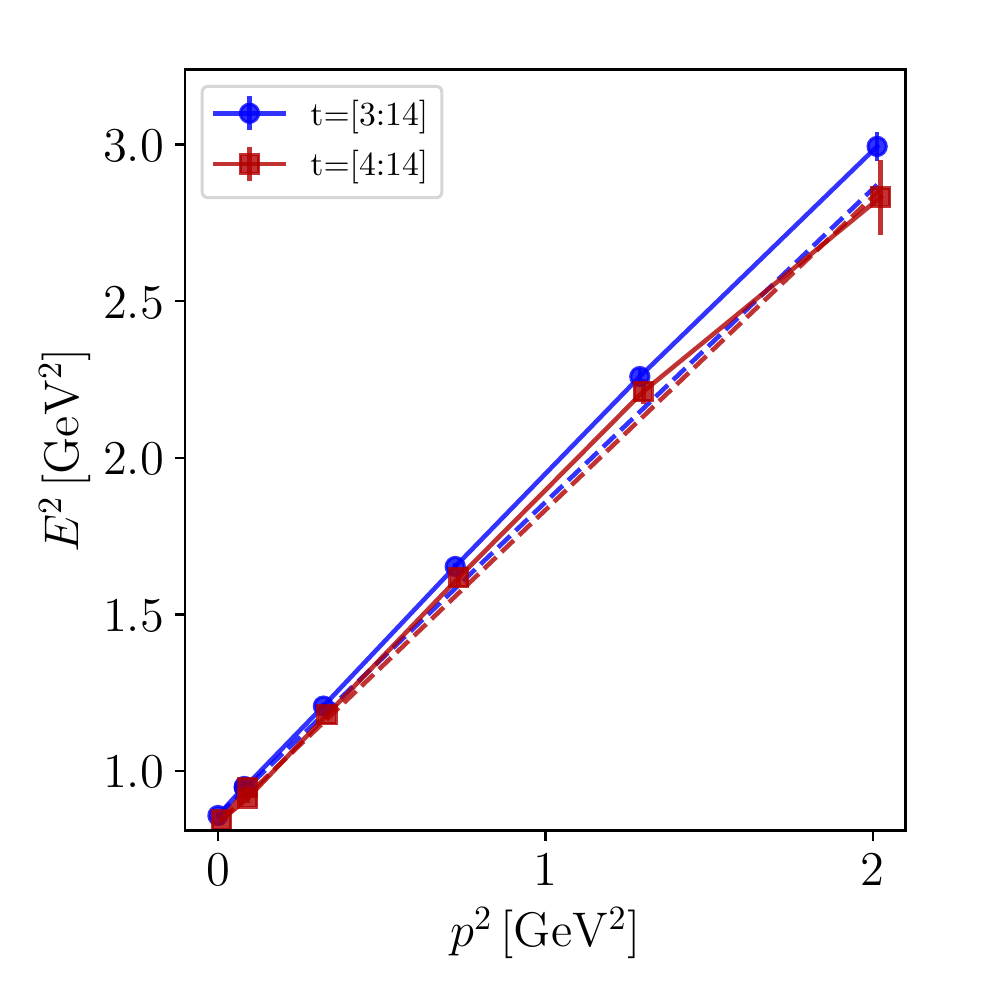}~
\includegraphics[width=.47\textwidth]{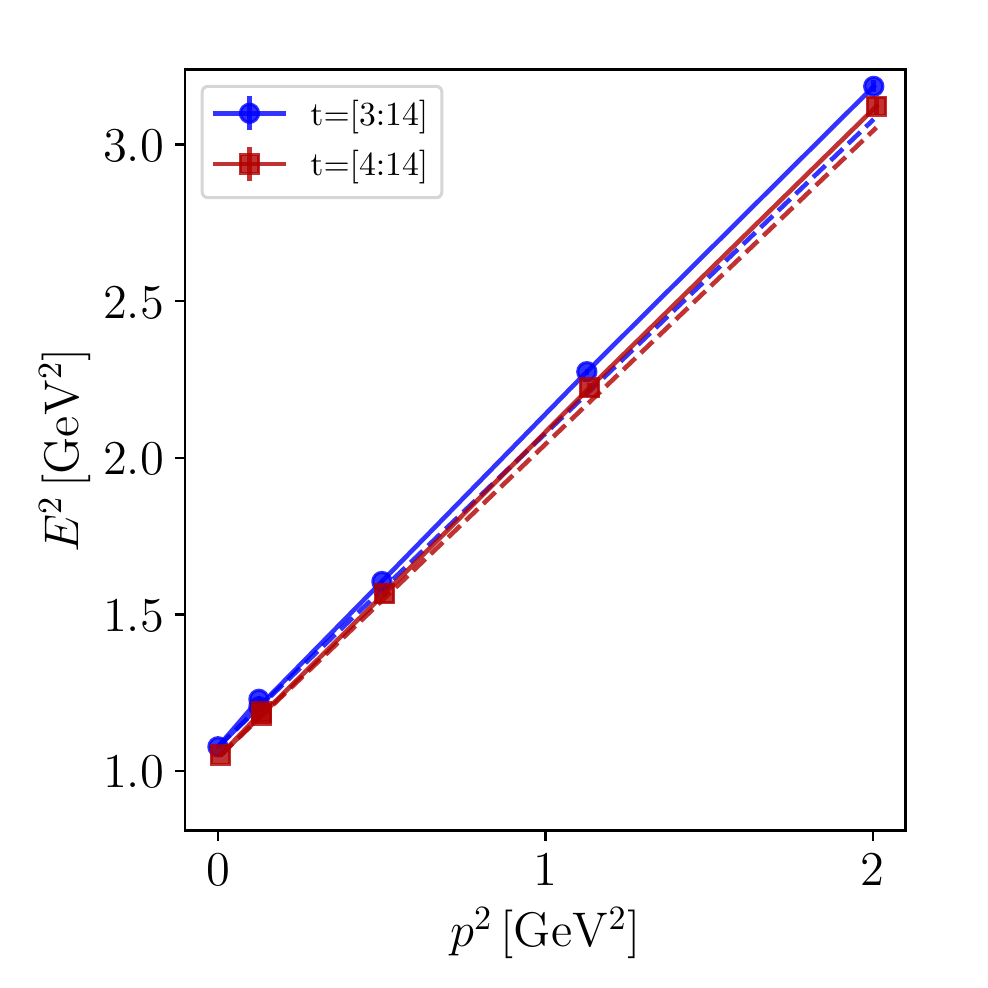}\\
\caption{The effective energy (top) and the dispersion relation of the nucleon (bottom)
determined on the D6 ($m_\pi\approx170\,\MeV$, $a\approx0.091\,\fm)$) (left)
and the E5 ($m_\pi\approx270\,\MeV$, $a\approx0.073\,\fm)$) (right) ensembles.
The dashed lines show the continuum dispersion relation with the mass determined from the
lattice.
  \label{fig:Eeff_disprel_cmp}}
\end{figure}

Nucleon matrix elements and form factors are extracted from nucleon-current three-point
correlation functions using standard methods of lattice QCD (see, e.g.
Ref.~\cite{Syritsyn:2009mx}).
We use the two-state model for both two- and three-point correlators,
\begin{align}
\label{eqn:c2fits}
\langle N(\vec p, t) \bar{N}(0)\rangle  &\sim C_0^2 e^{-E_{N0} t} + C_1^2 e^{-E_{N1} t}\,,\\ 
\label{eqn:c3fits}
\langle N(\vec p^\prime, t) J(\vec q, \tau) \bar{N}(0)  
  &\sim {\mathcal A}_{0^\prime 0} C_{0^\prime} C_0 e^{-E_{N0}^\prime(t-\tau) - E_{N0}\tau}
      + {\mathcal A}_{1^\prime 0} C_{1^\prime} C_0 e^{-E_{N1}^\prime(t-\tau) - E_{N0}\tau} \\
\nonumber
  &   + {\mathcal A}_{0^\prime 1} C_{0^\prime} C_1 e^{-E_{N0}^\prime(t-\tau) - E_{N1}\tau}
      + {\mathcal A}_{1^\prime 1} C_{1^\prime} C_1 e^{-E_{N1}^\prime(t-\tau) - E_{N1}\tau}
\end{align}
to find ground-state nucleon energies $E_{N0}^{(\prime)}$, matrix elements of
nucleon operators $C_{0^{(\prime)}}=\langle\mathrm{vac}|N|N(\vec p^{(\prime)})\rangle$,
and matrix elements of the vector current 
${\mathcal A}_{0^\prime 0}=\langle N(\vec p^\prime)|J|N(\vec p)\rangle$.
The two-point correlators and the ground-state energies are shown in
Fig.~\ref{fig:Eeff_disprel_cmp}.
The dispersion relation $E^2(p^2)$, also shown in Fig.~\ref{fig:Eeff_disprel_cmp}, is in good
agreement with the continuum $E^2(p)=E^2(0)+p^2$, indicating that discretization effects on the
spectrum of moving nucleons are insignificant.


\section{Connected contributions\label{sec:conn}}

\begin{figure}
\centering
\includegraphics[width=\textwidth]{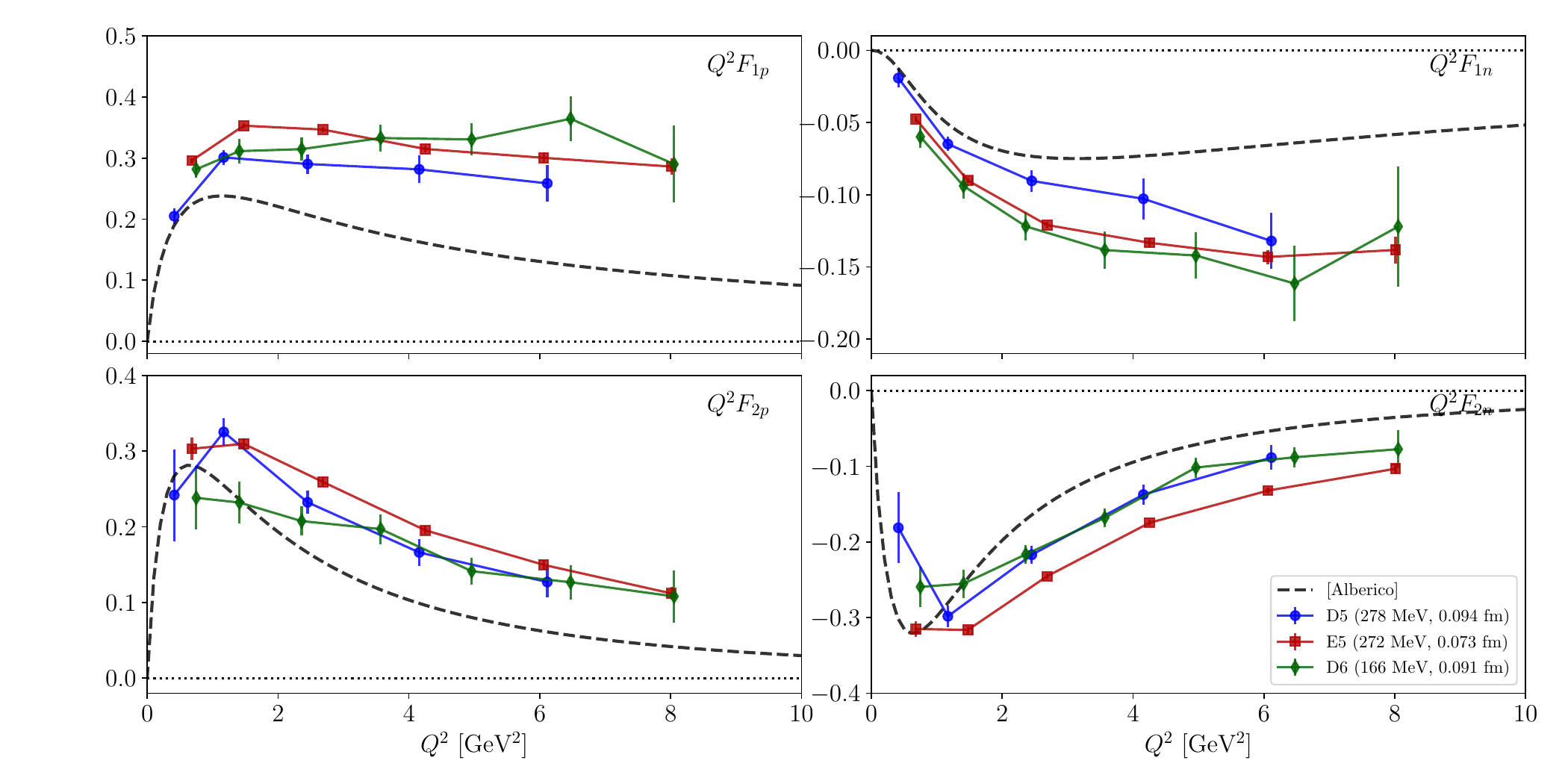}
\caption{Comparison of the Dirac $F_1$ (top) and the Pauli $F_2$ (bottom) form factors 
of the proton (left) and the neutron(right) computed on three ensembles D5, D6, and E5
(connected contributions only).
Ground-state matrix elements are extracted with 2-state fits to two- and three-point correlators
using data with source-sink separations $\tsep=0.7\ldots1.1\,\fm$.
The dashed lines show phenomenological fits~\cite{Alberico:2008sz}.
  \label{fig:cmp_F12_PN}}
\end{figure}

The ground-state matrix elements ${\mathcal A}^q_{0'0}$ from fits~(\ref{eqn:c3fits}) are decomposed
into form factors $F^q_{1,2}$~(\ref{eqn:formfacdef}) separately for flavors $u$ and $d$.
Light-flavor connected combinations corresponding to the proton and the neutron form factors are shown 
in Fig.~\ref{fig:cmp_F12_PN}, and compared to phenomenological fits~\cite{Alberico:2008sz}.
Although our lattice results have qualitatively similar $Q^2$ behavior to the phenomenological
fits, they overshoot the latter by a factor of $(2\ldots2.5)$.

This substantial difference may be due to any combination of excited state contributions, 
discretization effects, and omission of disconnected contributions.
Of these three potential sources, the excited state effects will likely be the most difficult to control.
Our calculations with two lattice spacings $a=0.091$ and $0.073\,\fm$ 
offer some insight into the magnitude of discretization effects.
In particular, the Dirac $F_1$ form factor appears less affected by them compared to the Pauli
$F_2$ form factor.
This difference can be attributed to how these form factors contribute to the vector current
matrix elements: unlike the Dirac $F_1$, the Pauli $F_2$ form factor contributes proportionally
to the momentum transfer $\propto\sigma_{\mu\nu}q^\nu$.
A detailed study of $O(a)$-improved current operators and calculations at additional values of
lattice spacings planned in the future will be highly beneficial for exploring these effects.

\section{Disconnected contributions\label{sec:disc}}
The main part of the present work is evaluating the magnitude of disconnected contributions to
the high-momentum nucleon form factors for the first time.
The nucleon-nucleon part of the Wick contractions in Eq.~(\ref{eqn:c3pt}) is evaluated
simultaneously with the connected 3-point correlators.
The ``quark loops'' 
\begin{equation}
\label{eqn:qloop}
\sum_{\vec z} e^{i\vec q\vec z} \la \bar q_z\Gamma q_z \ra_\text{Wick} 
= -\mathrm{Tr}\big[\Gamma \Dslash^{-1}\big]
\end{equation}
have to be calculated separately by (partially) stochastic methods.
To do that, we employ hierarchical probing combined with deflation of the
$(\Dslash^\dagger\Dslash)$ operator~\cite{Gambhir:2016jul,Gambhir:2016uwp}.
On each configuration, we calculate $N_{vec}=400$ low-lying eigenmodes that are used for (1)
exact evaluation of their contributions to Eq.~(\ref{eqn:qloop}) and (2) deflation of
hierarchical probing vectors~\cite{Gambhir:2016jul} that results in a substantial reduction of 
the variance in the stochastic estimation of the high-mode part.
We use $N_{hp}=512$ spatial hierarchical probing vectors on each configuration that result in
exact cancellation of variance contributions up to distance of $2a$ in each of the 4 directions.

\begin{figure}
\centering
\includegraphics[width=.38\textwidth]{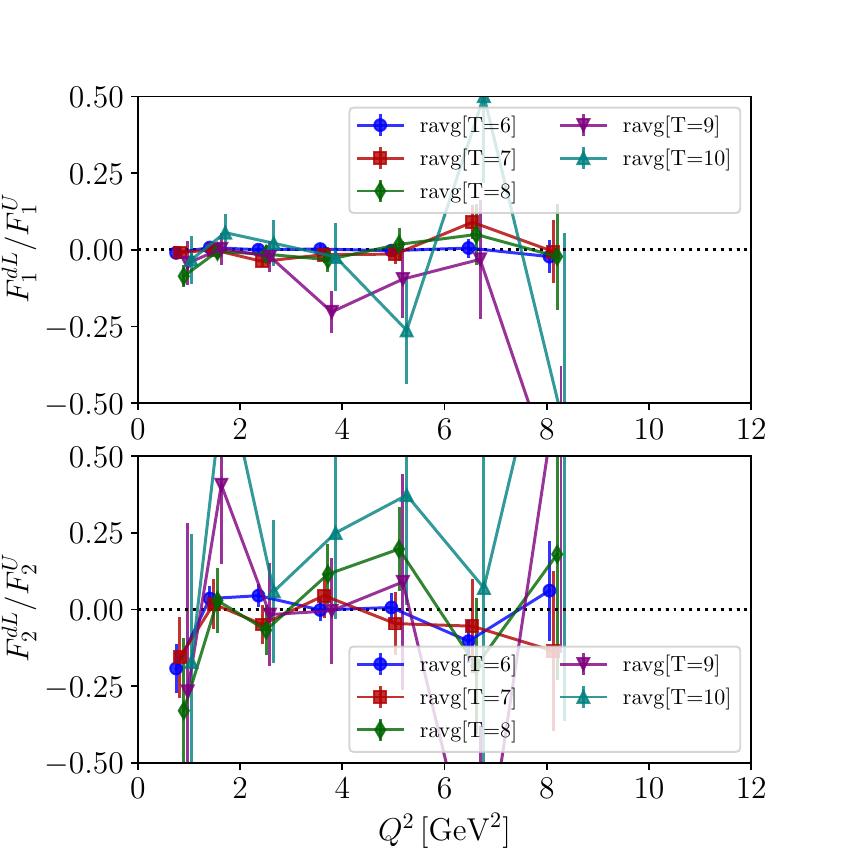}~
\hspace{-.05\textwidth}~
\includegraphics[width=.33\textwidth]{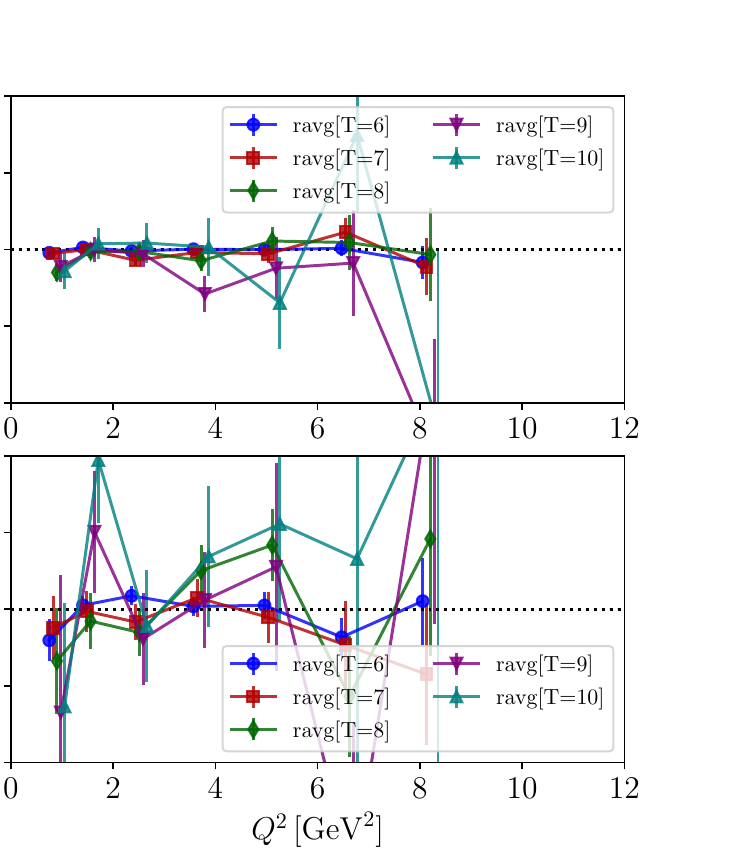}~
\hspace{-.05\textwidth}~
\includegraphics[width=.33\textwidth]{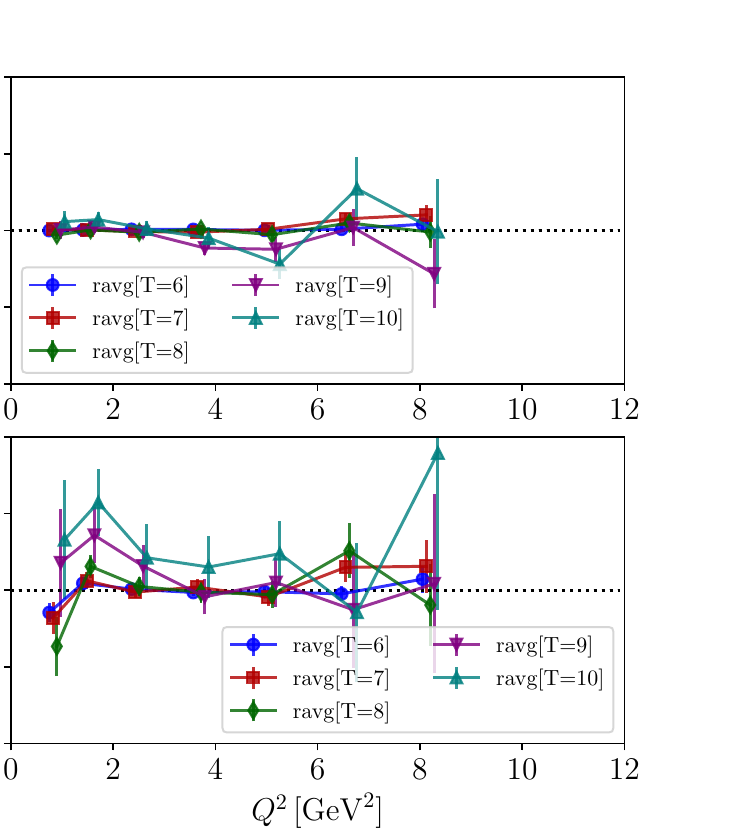}
\caption{Comparison of connected and disconnected contributions to the Dirac $F_1$ and Pauli
$F_2$ form factors of the proton on the D6 ($m_\pi\approx170\,\MeV$) ensemble: 
(left) light flavors $L$, (middle) strange flavor $S$ and 
(right) the combination $(L-S)$ occurring in the proton and the neutron. 
Note that statistical fluctuations cancel substantially in the latter.
  \label{fig:ratio_F12_disc2conn}}
\end{figure}

We have calculated disconnected quark loops and their contributions to the form factors on the
D5 and D6 ensembles, with full and $\approx1/2$ statistics respectively.
First we examine the magnitude of these disconnected contractions relative to the connected
contributions.
Since our current goal is to examine how much these disconnected contractions and their uncertainties
will affect the final result, we resort to using simple plateau-center averages as approximate estimators 
of the ground-state matrix elements.
In Figure~\ref{fig:ratio_F12_disc2conn}, we show the ratio of the disconnected to $u$
quark-connected contributions to the form factors $F_1$ and $F_2$ obtained on the D6 ensemble
with a range of source-sink separations $\tsep$.
We show separately the cases of disconnected light ($L$), strange ($S$), and the $[L-S]$
flavor combinations.
In all cases, the values are statistically compatible with zero.
However, the statistical uncertainties grow dramatically with increasing source-sink separation
$\tsep$, showing that it will be extremely challenging to constrain these contributions
especially if excited states play a significant role.
It is nevertheless reassuring that the $[L-S]$ flavor combination that contributes to the proton
($P$) and neutron ($N$) form factors,
\begin{equation}
\begin{aligned}
P &= \frac13[2U - D]_\text{conn} + \frac13[L-S]_\text{disc}\,,\\
N &= \frac13[2D - U]_\text{conn} + \frac13[L-S]_\text{disc}\,,
\end{aligned}
\end{equation}
has the smallest statistical uncertainties due to partial noise cancellation.
In particular, we expect that statistical uncertainty in $F_1$ due to disconnected contributions
will be $\lesssim20\%$ for $Q^2$ up to $8\,\GeV^2$, 
unlike in individual flavor form factors $F_{1,2}^{u,d,s}$, which will be more challenging to
compute with high precision.

\begin{figure}
\centering
\includegraphics[width=.47\textwidth]{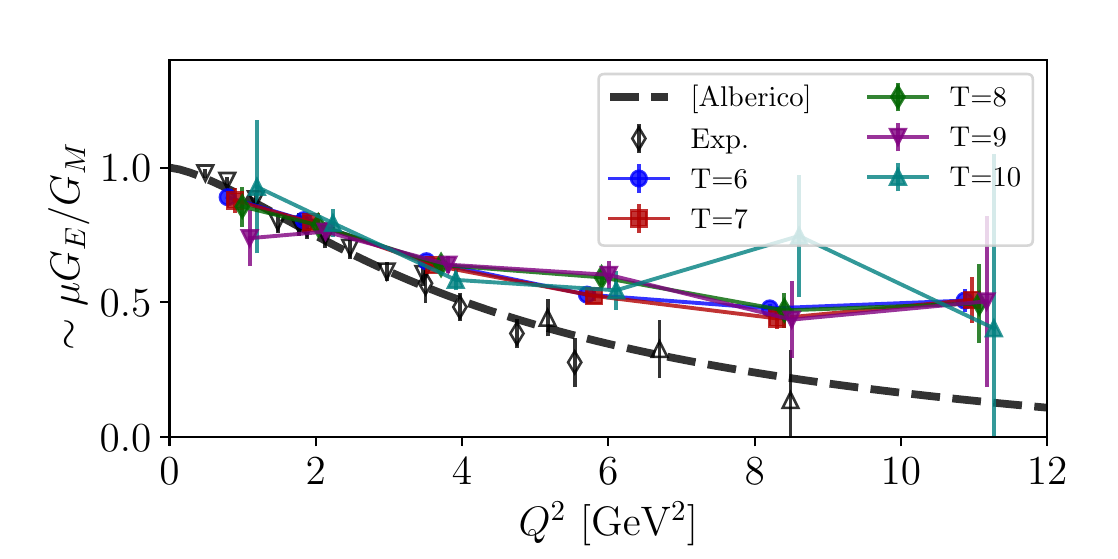}~
\includegraphics[width=.47\textwidth]{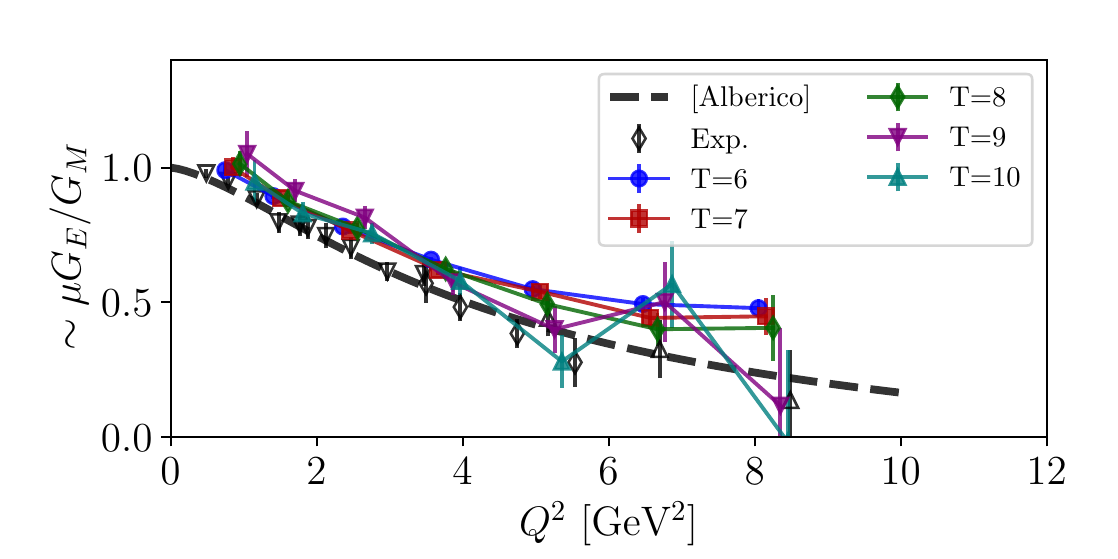}\\
\includegraphics[width=.47\textwidth]{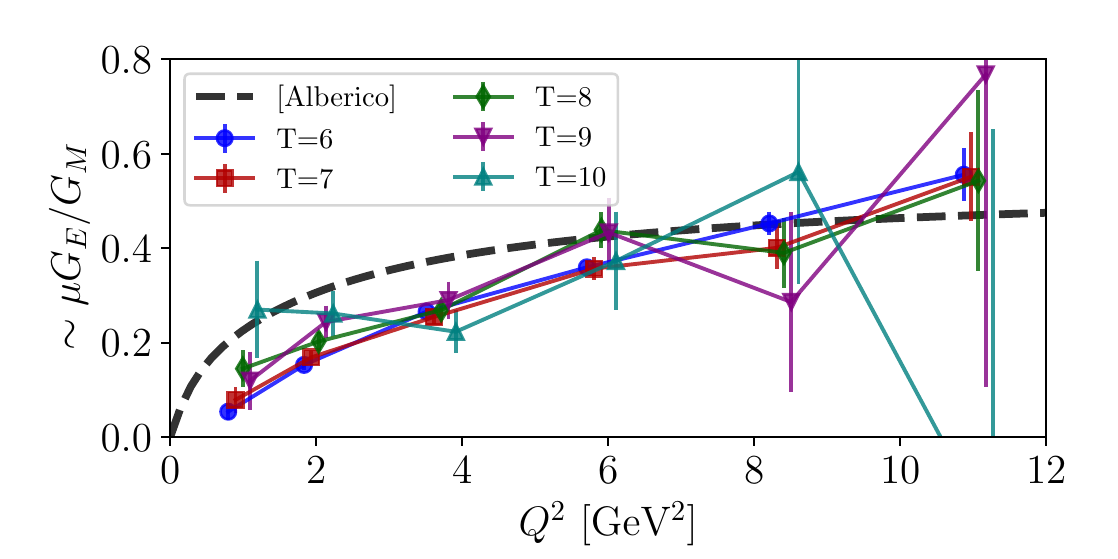}~
\includegraphics[width=.47\textwidth]{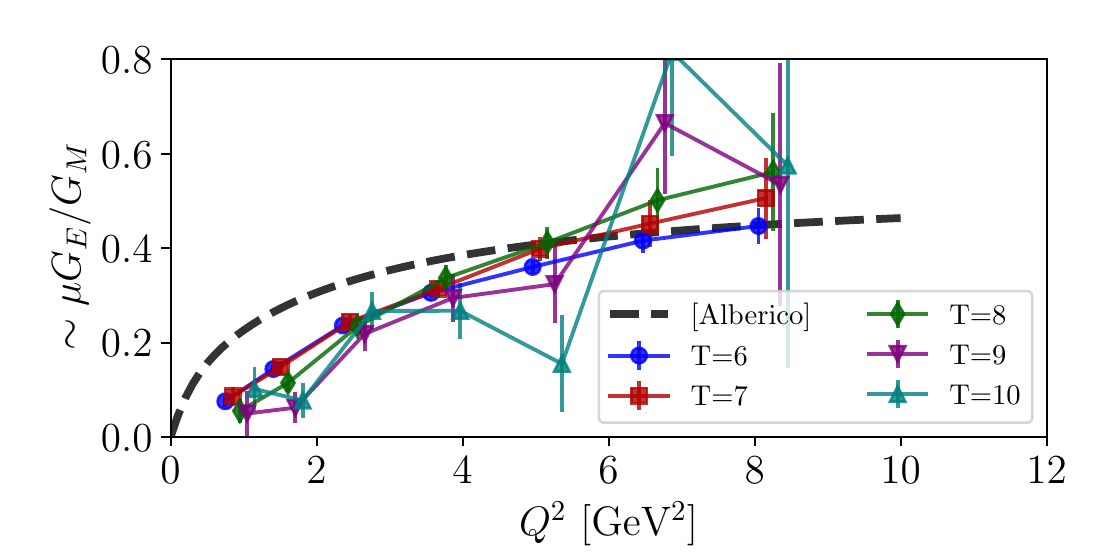}\\
\caption{Ratio of the form factors $G_E/G_M$ of the proton (top) and neutron (bottom)
on the D5 ($m_\pi\approx280\,\MeV$, left) and D6 ($m_\pi\approx170\,\MeV$, right) ensembles.
Both connected and disconnected contributions are included.
The black data points are experimental values and the dashed lines are phenomenological
fits~\cite{Alberico:2008sz}
  \label{fig:cmp_GE2GM_PN}}
\end{figure}

Finally, in Figure~\ref{fig:cmp_GE2GM_PN} we show our preliminary results for the 
proton and the neutron form factor ratios $G_E/G_M$ with both connected and disconnected
contributions included.
As in our previous calculations, the ratios are much closer to the phenomenological fits despite
the disagreement in the individual form factors.
On the D6 ensemble, which is the closest one to the physical point, we observe qualitative agreement
with phenomenology, and the statistical uncertainty of our plateau-average result at
$\tsep\approx0.7\ldots0.8\,\fm$ is comparable to previous experiments.
One notable exception is the $G_E/G_M$ ratio for the neutron at small $Q^2\lesssim2\,\GeV^2$.
This disagreement is likely attributable to excited state contamination.
Indeed, since our momentum-smeared nucleon operators are optimized for overlap with the ground
state in the Breit frame, they become poorly optimized when the source momentum is near
zero, i.e. $Q^2\lesssim \frac14Q^2_\text{max}$.

\section{Summary/outlook}

In this report, we present our preliminary results from high-statistics calculations of the 
proton and neutron electromagnetic form factors at high momentum up to $Q^2\approx 8\GeV^2$.
We evaluate both connected and disconnected quark contributions at two different lattice
spacings and two pion masses, one of which is very close to the physical point.
Our preliminary analysis indicates that stochastic uncertainty due to disconnected
contributions can be constrained to $\lesssim20\%$ for the proton and neutron form factors but
it will be more challenging in the cases of individual flavors.
Our current results pave the way for a complete calculation of high-momentum form factors that
will be compared to the new data obtained at JLab.

\acknowledgments
S.S. is supported by the National Science Foundation under NSF award PHY 2412963.
Any opinions, findings, and conclusions or recommendations expressed in this material are those
of the author(s) and do not necessarily reflect the views of the National Science Foundation.
M.E., J.N. and A.P. are supported by the U.S. DOE, Office of Science,
Office of Nuclear Physics through grants numbered DE-FG02-96ER40965, DE-SC-0011090 and
DE-SC-0023116, respectively.
S.K. is supported by the MKW NRW under funding code NW21-024-A.  
The research reported in this work made use of computing and long-term storage facilities of the
USQCD Collaboration, which are funded by the Office of Science of the U.S. Department of Energy.
The authors gratefully acknowledge the Gauss Centre for Supercomputing e.V. (www.gauss-centre.eu)
for funding this project by providing computing time through the John von Neumann Institute for
Computing (NIC) on the GCS Supercomputer JUWELS at J\"ulich Supercomputing Centre (JSC).
We are grateful to our JLab/W\&M, LANL, and MIT colleagues for supplying dynamical Wilson-clover 
gauge configurations for this study.
This research also used resources of the National Energy Research Scientific Computing Center, a DOE
Office of Science User Facility supported by the Office of Science of the U.S. Department of Energy
under Contract No. DE-AC02-05CH11231 using NERSC award NP-ERCAP0024043.
The computations were performed using the Qlua software suite~\cite{qlua-software}.




\begin{thebibliography}{10}

\bibitem{Punjabi:2015bba}
V.~Punjabi, C.F.~Perdrisat, M.K.~Jones, E.J.~Brash and C.E.~Carlson, \emph{{The
  Structure of the Nucleon: Elastic Electromagnetic Form Factors}},
  \href{https://doi.org/10.1140/epja/i2015-15079-x}{\emph{Eur. Phys. J.}
  {\bfseries A51} (2015) 79}
  [\href{https://arxiv.org/abs/1503.01452}{{\ttfamily 1503.01452}}].

\bibitem{Cui:2020rmu}
Z.-F.~Cui, C.~Chen, D.~Binosi, F.~de~Soto, C.D.~Roberts,
  J.~Rodr\'\i{}guez-Quintero et~al., \emph{{Nucleon elastic form factors at
  accessible large spacelike momenta}},
  \href{https://doi.org/10.1103/PhysRevD.102.014043}{\emph{Phys. Rev. D}
  {\bfseries 102} (2020) 014043}
  [\href{https://arxiv.org/abs/2003.11655}{{\ttfamily 2003.11655}}].

\bibitem{jlabHallA_HRS_gmp17GeV}
J.~Arrington, E.~Christy, S.~Gilad, B.~Moffit, V.~Sulkosky, B.~Wojtsekhowski et~al., 
\emph{Precision Measurement of the Proton Elastic Cross Section at High  $Q^2$}. 
\href{https://www.jlab.org/exp_prog/proposals/07/PR12-07-108.pdf}{DOI},
2007.

\bibitem{jlabHallA_SBS_gep15GeV}
E.~Brash, E.~Cisbani, M.~Jones, M.~Khandaker, N.~Liyanage, L.~Pentchev et~al.,
\emph{Large Acceptance Proton Form Factor Ratio Measurements at 13 and 15 $(\mathrm{GeV}/c)^2$ 
Using Recoil Polarization Method}.
\href{https://www.jlab.org/exp_prog/proposals/07/PR12-07-109.pdf}{DOI}, 
2008.

\bibitem{jlabHallA_SBS_gen10GeV}
G.~Cates, S.~Riordan, B.~Wojtsekhowski et~al., 
\emph{Measurement of the Neutron Electromagnetic Form Factor Ratio $G_{En}/G_{Mn}$ at High $Q^2$}.
\href{https://www.jlab.org/exp_prog/proposals/09/PR12-09-016.pdf}{DOI}, 
2009.

\bibitem{jlabHallA_SBS_gmn18GeV}
J.~Annand, R.~Gilman, B.~Quinn, B.~Wojtsekhowski et~al., 
\emph{Precision measurement of the Neutron Magnetic Form Factor up to $Q^2=18.0
(\mathrm{GeV}/c)^2$ by the Ratio Method}.
\href{https://www.jlab.org/exp_prog/proposals/09/PR12-09-019.pdf}{DOI}, 
2009.

\bibitem{jlabHallB_CLAS12_gmn14GeV}
W.~Brooks, J.~Lachniet, M.~Vineyard et~al., 
\emph{Measurement of the Neutron Magnetic Form Factor at High $Q^2$ Using the Ratio Method on Deuterium}.
\href{https://www.jlab.org/exp_prog/proposals/07/PR12-07-104.pdf}{DOI}, 
2007.

\bibitem{Christy:2021snt}
M.E.~Christy et~al., \emph{{Form Factors and Two-Photon Exchange in High-Energy
  Elastic Electron-Proton Scattering}},
  \href{https://doi.org/10.1103/PhysRevLett.128.102002}{\emph{Phys. Rev. Lett.}
  {\bfseries 128} (2022) 102002}
  [\href{https://arxiv.org/abs/2103.01842}{{\ttfamily 2103.01842}}].

\bibitem{Chambers:2017tuf}
{\scshape UKQCD, QCDSF, CSSM} collaboration, \emph{Electromagnetic form
  factors at large momenta from lattice QCD},
  \href{https://doi.org/10.1103/PhysRevD.96.114509}{\emph{Phys. Rev.}
  {\bfseries D96} (2017) 114509}
  [\href{https://arxiv.org/abs/1702.01513}{{\ttfamily 1702.01513}}].

\bibitem{Lepage:1989hd}
G.P.~Lepage, \emph{The analysis of algorithms for lattice field theory}. 
Invited lectures given at TASI'89 Summer School, Boulder, CO, Jun 4-30, 1989.

\bibitem{Syritsyn:2017jrc}
S.~Syritsyn, A.S.~Gambhir, B.~Musch and K.~Orginos, \emph{{Constructing Nucleon
  Operators on a Lattice for Form Factors with High Momentum Transfer}},
  \href{https://doi.org/10.22323/1.256.0176}{\emph{PoS} {\bfseries LATTICE2016} (2017) 176}.

\bibitem{Kallidonis:2018cas}
C.~Kallidonis, S.~Syritsyn, M.~Engelhardt, J.~Green, S.~Meinel, J.~Negele
  et~al., \emph{{Nucleon electromagnetic form factors at high $Q^2$ from
  Wilson-clover fermions}},
  \href{https://doi.org/10.22323/1.334.0125}{\emph{PoS} {\bfseries LATTICE2018}
  (2018) 125} [\href{https://arxiv.org/abs/1810.04294}{{\ttfamily
  1810.04294}}].

\bibitem{Syritsyn:2023pmn}
S.~Syritsyn, M.~Engelhardt, J.~Green, S.~Krieg, J.~Negele and A.~Pochinsky,
  \emph{{Nucleon Electromagnetic Form Factors at Large Momentum Transfer from
  Lattice QCD}}, \href{https://doi.org/10.1007/s00601-023-01839-4}{\emph{Few
  Body Syst.} {\bfseries 64} (2023) 72}.

\bibitem{Bali:2016lva}
G.S.~Bali, B.~Lang, B.U.~Musch and A.~Sch{\"a}fer, \emph{{Novel quark smearing
  for hadrons with high momenta in lattice QCD}},
  \href{https://doi.org/10.1103/PhysRevD.93.094515}{\emph{Phys. Rev.}
  {\bfseries D93} (2016) 094515}
  [\href{https://arxiv.org/abs/1602.05525}{{\ttfamily 1602.05525}}].

\bibitem{Green:2015wqa}
J.~Green, S.~Meinel, M.~Engelhardt, S.~Krieg, J.~Laeuchli, J.~Negele et~al.,
  \emph{{High-precision calculation of the strange nucleon electromagnetic form
  factors}}, \href{https://doi.org/10.1103/PhysRevD.92.031501}{\emph{Phys.
  Rev.} {\bfseries D92} (2015) 031501}
  [\href{https://arxiv.org/abs/1505.01803}{{\ttfamily 1505.01803}}].

\bibitem{Syritsyn:2009mx}
S.~Syritsyn, J.~Bratt, M.~Lin, H.~Meyer, J.~Negele et~al., \emph{{Nucleon
  Electromagnetic Form Factors from Lattice QCD using 2+1 Flavor Domain Wall
  Fermions on Fine Lattices and Chiral Perturbation Theory}},
  \href{https://doi.org/10.1103/PhysRevD.81.034507}{\emph{Phys. Rev.} {\bfseries
  D81} (2010) 034507} [\href{https://arxiv.org/abs/0907.4194}{{\ttfamily
  0907.4194}}].

\bibitem{Alberico:2008sz}
W.M.~Alberico, S.M.~Bilenky, C.~Giunti and K.M.~Graczyk, \emph{{Electromagnetic
  form factors of the nucleon: New Fit and analysis of uncertainties}},
  \href{https://doi.org/10.1103/PhysRevC.79.065204}{\emph{Phys. Rev.}
  {\bfseries C79} (2009) 065204}
  [\href{https://arxiv.org/abs/0812.3539}{{\ttfamily 0812.3539}}].

\bibitem{Gambhir:2016jul}
A.S.~Gambhir, A.~Stathopoulos, K.~Orginos, B.~Yoon, R.~Gupta and S.~Syritsyn,
  \emph{{Algorithms for Disconnected Diagrams in Lattice QCD}}, 
\href{https://doi.org/10.22323/1.256.0265}{\emph{PoS}
  {\bfseries LATTICE2016} (2016) 265}
  [\href{https://arxiv.org/abs/1611.01193}{{\ttfamily 1611.01193}}].

\bibitem{Gambhir:2016uwp}
A.S.~Gambhir, A.~Stathopoulos and K.~Orginos, \emph{{Deflation as a Method of
  Variance Reduction for Estimating the Trace of a Matrix Inverse}},
  \href{https://doi.org/10.1137/16M1066361}{\emph{SIAM J. Sci. Comput.}
  {\bfseries 39} (2017) A532}
  [\href{https://arxiv.org/abs/1603.05988}{{\ttfamily 1603.05988}}].

\bibitem{qlua-software}
A.~Pochinsky, \emph{Qlua lattice software suite.}
  \href{https://usqcd.lns.mit.edu/qlua}{DOI}, 2008--present.

\end{thebibliography}

\providecommand{\href}[2]{#2}\begingroup\raggedright\endgroup

\end{document}